\documentclass[onecolumn,usenatbib]{mn2e}

\usepackage{amsmath,amssymb}
\usepackage{bm}
\usepackage{cases}

\def\be{\begin{equation}}
\def\ee{\end{equation}}
\def\ba{\begin{eqnarray}}
\def\ea{\end{eqnarray}}
\def\nn{\nonumber}

\newcommand{\eref}[1]{Eq.~(\ref{#1})}

\usepackage{color}              
\usepackage{graphicx}   
\usepackage{comment}
\usepackage{hyperref}
\hypersetup{
    colorlinks=true,
    linkcolor=red,
    citecolor=blue,
}

\newcommand\nat{Nature}
\newcommand\apjl{Astrophys.~J.~Lett.}
\newcommand\apj{Astrophys.~J.}
\newcommand\prd{Phys.~Rev.~D}
\newcommand\jcap{JCAP}
\newcommand\qjras{Royal~Astronomical~Society,~Quarterly~Journal}
\newcommand\mnras{Month.~Not.~R.~Astro.~Soc.}
\newcommand\prl{Phys.~Rev.~Lett.}

\begin{document}

\title{
Parity-odd correlators of diffuse gamma rays and intergalactic magnetic fields
}

\date{\today}

\author[H. Tashiro \& T. Vachaspati]{Hiroyuki Tashiro$^*$ and Tanmay
Vachaspati$^\dag$\\
$^*$Department of Physics and Astrophysics, Nagoya University, Nagoya 464-8602, Japan. \\
$^\dag$Physics Department, Arizona State University, Tempe, Arizona 85287, USA.
}

\maketitle

\begin{abstract}
We develop the connection between intergalactic helical magnetic fields 
and parity odd signatures in the diffuse gamma ray sky. 
We find that the location and the amplitude of a peak in a parity odd correlator, $Q(R)$, can be 
used to infer the normal and helical power spectra of the intergalactic magnetic field.
When applied to Fermi-LAT data, the amplitude of the observed peak in $Q(R)$ gives
$\sim 10^{-14}~{\rm G}$ intergalactic magnetic field strength, which is consistent with an
earlier independent estimate that only used the peak location~\citep{Tashiro:2013ita}.
We discuss features in the observed $Q(R)$ that further support the intergalactic magnetic field 
hypothesis and make predictions for future tests.
\end{abstract}

\section{Introduction}
\label{introduction}

It has been known for some time that gamma rays from TeV blazars can probe the intergalactic 
magnetic field~\citep{Aharonian:1993vz,Plaga:1995,ner,Elyiv:2009bx,Dolag:2009iv,Neronov:2009gh}.
Photons with TeV energy from such beamed sources scatter off the 
extra-galactic background light (EBL) and produce electron-positron pairs also with 
TeV energy~\citep{gould,Aharonian:2005gh}.
These charged particles then inverse-Compton scatter off cosmic microwave background (CMB) photons
and up-scatter them to a cascade of GeV photons. These GeV gamma rays potentially carry 
information about the electron-positron trajectories. In particular, if the electron-positron are deflected
by ambient magnetic fields, the cascade gamma rays observed at GeV energies also carry information 
about the ambient magnetic fields. 

A few years ago, observations of four TeV blazars by the Fermi Large
Area Telescope~(Fermi-LAT) experiment\footnote{http://fermi.gsfc.nasa.gov}
and the High Energy
Stereoscopic System (HESS) gamma ray telescope\footnote{http://www.mpi-hd.mpg.de/hfm/HESS/} were used to obtain a lower bound
of $\sim 10^{-16}~{\rm G}$ on the intergalactic magnetic field strength~\citep{Neronov:1900zz,Tavecchio:2010mk,Dolag:2010ni}. The conclusion
is based on the observed deficit of GeV photons, the assumption being that the deficit is due to
deflection of the electron-positron pairs due to an inter-galactic magnetic field.
The assumption has since been debated~\citep{Broderick:2011av,
Miniati:2012ge,2012ApJ...758..102S}, with the concern that there is an instability in 
the propagation of high energy charged particles in the cosmological medium that tends to isotropize 
the directions of the charged particles thus explaining the observed deficit of GeV photons. 

More recently, developing ideas first proposed in the context of cosmic rays~\citep{Kahniashvili:2005yp},
we have shown that the cascade GeV gamma rays from TeV blazars may also be
used to probe {\em helical} magnetic fields as the helicity introduces a parity odd signature
in the arrival directions of the gamma rays~\citep{Tashiro:2013bxa}. If observed, a parity violating
signal around blazars would be hard to explain on the basis of a plasma instability but easier to 
explain with magnetic field helicity.
In turn, primordial magnetic field helicity would be an invaluable probe of CP violation in the
early universe with profound consequences for very high energy particle physics and the origin 
of matter~\citep{Vachaspati:2001nb,Copi:2008he,Chu:2011tx,Long:2013tha}.
A magnetic field in the post-recombination universe can be a critical ingredient that enables
 cosmic structure formation~\citep{Rees:1987}. 

Most recently, we have extended the scheme to look for parity odd signatures in gamma
rays from a single source to the diffuse gamma ray background over the entire sky~\citep{Tashiro:2013ita}. 
This extension is particularly necessary if the cascade deflections are large and the observed 
GeV gamma rays are not obviously associated with any TeV blazar source. Further, we have
applied our scheme to the diffuse gamma rays observed by Fermi-LAT, and find a significant
parity odd signal. Interpreted in terms of a helical magnetic field, we find a field strength
$\sim 10^{-14}~{\rm G}$ on intergalactic scales with left-handed helicity.

If ongoing observations continue to confirm the parity odd signal, we should be able to use the 
gamma ray signal to reconstruct spectral properties of 
the intergalactic field. More specifically, the equal-time correlation function of a stochastic, isotropic, 
magnetic field in Minkowski spacetime
can be written as~\citep{monin}
\be
\langle B_i ({\bm x}) B_j ({\bm y}) \rangle = M_N (r) \left ( \delta_{ij} - \frac{r_i r_j}{r^2} \right )
       + M_L(r) \frac{r_i r_j}{r^2} + M_H (r) \epsilon_{ijl} {r^l},
\label{magcorr}
\ee
where ${\bm r} = {\bm x}-{\bm y}$ and, due to the divergence free condition of magnetic
fields,
\begin{equation}
 M_N(r) ={1 \over 2 r} {d \over dr} (r^2 M_L(r)).
\end{equation}
We would like to relate the ``normal'' and ``helical'' power spectra, $M_N$ and $M_H$, to correlators of the 
observed cascade GeV gamma rays. 
In~\citet{Tashiro:2013ita}, we had related $M_H(r)$ to a parity odd correlator of cascade gamma rays from
a known source. In the present paper we will extend that analysis to the case of diffuse gamma
rays, when the source locations are not known. Further, if $M_H \ne 0$, it allows for a new way to
also estimate the normal power spectrum from correlation functions of cascade gamma rays.

In Sec.~\ref{parityoddcorrelators} we will introduce our strategy in more detail, first in the context of a
single source, then in the context of many sources and when additional non-cascade photons
are present in the data. We will find that the parity odd signal can be a valuable tool for extracting
information not just about the magnetic field but also about the relative number of cascade
and non-cascade photons, which is related to the number of TeV blazars. In Sec.~\ref{samplemodels}
we make predictions for the parity odd signal if the magnetic field correlator has a simple power
law form. We conclude in Sec.~\ref{conclusions} where we also discuss limitations
of the present analysis. In Appendix~\ref{sec:sign} we relate the sign of the parity odd statistic
to the handedness of the magnetic field.

\section{Parity odd correlators of gamma rays}
\label{parityoddcorrelators}

\begin{figure}
  \begin{center}
  \includegraphics[width=150mm]{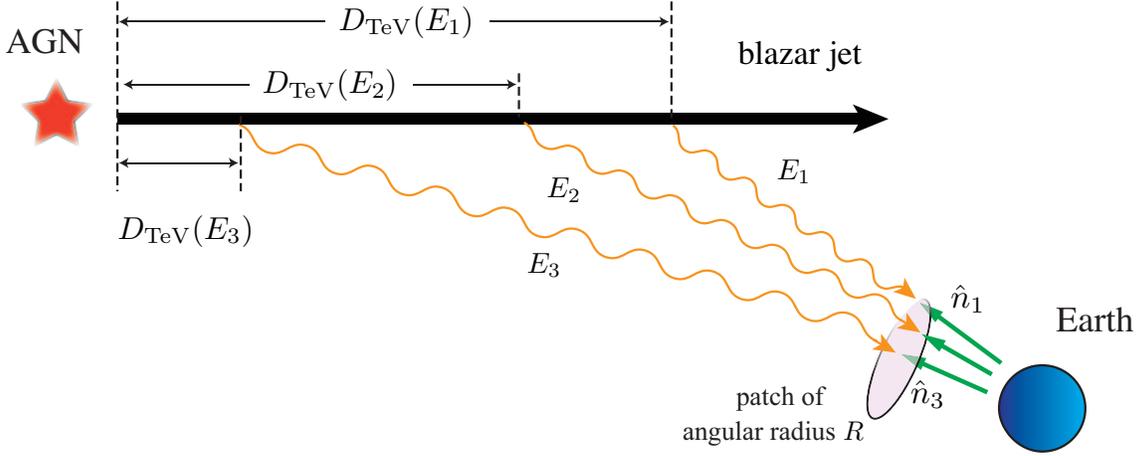}
  \end{center}
\caption{A collimated jet emanates from the source AGN. Very high energy photons 
propagate a short distance, $D_{\rm TeV}(E_3)$, and then pair-produce. The charged pairs
propagate a very short distance (too small to be shown) and then inverse Compton scatter
CMB photons to GeV energies
that are then observed to come from a direction ${\hat {\bm n}}_3$. Lower energy photons
from the source propagate a larger distance, $D_{\rm TeV}(E_1)$, before producing pairs, that
then produce cascade photons seen to come from direction ${\hat {\bm n}}_1$. 
The statistics, $Q(R)$, is the average of the triple product of the vectors ${\hat {\bm n}}_1$,
${\hat {\bm n}}_2$ (not labeled in the diagram) and ${\hat {\bm n}}_3$ in a patch of radius $R$
where $E_3 > E_2 > E_1$.
 }
\label{bendingschematics}
\end{figure}

Consider a TeV blazar at redshift $z_s$. The emitted TeV gamma rays interact with the EBL
to produce electron-positron pairs over a mean free path~\citep{Neronov:2009gh}
(see Fig.~\ref{bendingschematics})
\begin{equation}
 D_{\rm TeV} (E_{\rm TeV}) \sim 80 { \kappa \over (1+z_s)^2}  ~{\rm Mpc}~\left( {E_{\rm TeV} \over 10~{\rm TeV}}\right)^{-1},
 \label{eq:tev-meanfree}
\end{equation}
where $\kappa$ is a numerical factor which accounts
for the model uncertainties of EBL. Here we take $\kappa \sim 1$~\citep{Neronov:2009gh}.

Electron-positron pairs generated by the TeV gamma ray lose energy by IC scattering off CMB photons
over a distance~\citep{Neronov:2009gh}
\begin{equation}
 D_e \sim 
 30 ~{\rm kpc} ~ (1+z_e)^{-4} \left( {E_e \over 10~{\rm TeV} }\right)^{-1},
 \label{Deeq}
\end{equation}
where $z_e$ is the typical redshift at which TeV gamma rays create pairs, and $E_e  \sim E_{\rm TeV}/2$ 
is the electron energy. The secondary gamma ray cascade contains up-scattered CMB photons with
typical energy
\begin{equation}
 E_\gamma = {4 \over 3} (1+z_e)^{-1} \epsilon_{CMB} \left( {E_e \over m_e }\right)^2
  \sim 88 ~{\rm GeV} ~\left( {E_{\rm TeV} \over 10~{\rm TeV} }\right)^{2},
  \label{eq:gamma-ray}
\end{equation}
where $\epsilon_{\rm CMB}= 6 \times 10^{-4} (1+z_e)~{\rm eV}$ is the typical energy 
of CMB photons. 

The angle by which gamma rays are bent with respect to the source direction is given by~\citep{Tashiro:2013bxa}
\begin{equation}
 \Theta (E_\gamma) \approx
  {q D_{\rm TeV}  D_e \over E_e D_s}  ~{v}_{L} {B}
 \approx
 7.3 \times 10^{-5} ~ \left( { B_0  \over 10^{-16} ~{\rm G}}\right) \left( {E_\gamma
  \over 100 ~{\rm GeV}}\right)^{-3/2} \left( {D_s
  \over 1000 ~{\rm Mpc}}\right)^{-1} \left({1+z_s}\right)^{-4},
\label{eq:final_ang_est}
\end{equation}
where $q$ is the electron charge and $v_L \sim 1$ is the speed. To obtain
the redshift dependence we take the redshifts of the pair production event, $z_e$, to be
approximately the redshift to the source, $z_s$, and $B_0 = B/(1+z_e)^2$ as
expected from flux conservation.

In the following subsection, we will consider cascade photons of three different energies 
$E_1 < E_2 < E_3$ with arrival directions ${\bm n}(E_1)$, ${\bm n}(E_2)$
and ${\bm n}(E_3)$~(see Fig.~\ref{bendingschematics})
and evaluate the correlator
\be
Q_\infty = \langle {\bm n} (E_1) \times {\bm n} (E_2) \cdot {\bm n}(E_3) \rangle,
\label{Qinfty}
\ee
for a single source.
This is a special case of the correlator for diffuse background of gamma rays that was 
considered in~\citep{Tashiro:2013ita}
\be
Q(R) = \langle {\bm n} (E_1) \times {\bm n} (E_2) \cdot {\bm n}(E_3) \rangle_R,
\label{QR}
\ee
in which only photons within an angular distance $R$ from ${\bm n}(E_3)$ were included
in the ensemble average and then an average over all locations of ${\bm n}(E_3)$ was
taken. In other words, the ensemble average was taken over
photons such that ${\bm n}(E_1) \cdot {\bm n}(E_3) > \cos R$ and
${\bm n}(E_2) \cdot {\bm n}(E_3) > \cos R$ for all locations, ${\bm n}(E_3)$, on the sky.

\subsection{A single unidentified source}
\label{singlesource}

\begin{figure}
  \begin{center}
     \includegraphics[width=80mm]{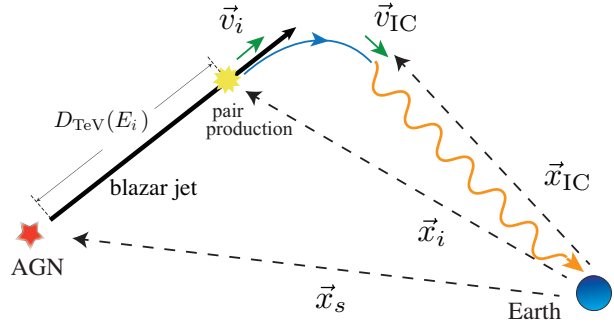}
  \end{center}
\caption{The TeV gamma ray from a blazar pair produces at ${\bm x}_i$ and the pair
has velocity ${\bm v}_i$. The pair bend in the magnetic field (only the $e^+$ trajectory
is shown) and then inverse Compton at ${\bm x}_{IC}$ when the velocity is ${\bm v}_{IC}$.
The constraint is that the up-scattered CMB photon, now at GeV energy and propagating
along ${\bm v}_{IC}$ reaches Earth. This gives the constraint equation: 
${\bm v}_{IC} = - {\hat {\bm x}}_{IC}$, where we choose the origin of the coordinate system
to be located at Earth.
}
\label{fig:schematics}
\end{figure}

We imagine a single beamed source of TeV gamma rays that is positioned in a general way --
the Earth may or may not lie within the opening angle of the jet.
Even if Earth is located outside the jet, cascade photons from 
the source can still get to Earth if intergalactic magnetic fields are present because the trajectories of the 
electron-positron pairs will bend in the magnetic field as illustrated in Fig.~\ref{fig:schematics}. 
We will consider this situation.

Let us denote the location of the source by ${\bm x}_s$. A $\sim {\rm TeV}$ energy photon
propagates a distance $D_{\rm TeV}$ with velocity ${\bm v}_i$ and then pair produces an
electron-positron pair at ${\bm x}_i$, at time $t_i$, given by
\be
{\bm x}_i = {\bm x}_s + D_{\rm TeV} {\bm v}_i.
\label{xiconstraint}
\ee
The electron-positron then propagate in a weak magnetic field which causes their trajectories
to bend. However, the bending is weak, and for the up-scattered CMB photons to arrive
at Earth, the velocity of the positron (say) at time $t_i$ has to approximately point to the Earth. 
Hence at time $t_i$ we can write,
\be
{\bm v}_i = - {\hat {\bm x}}_i + \delta {\bm v}_i,
\label{vieq}
\ee
with ${\hat {\bm x}}_i \cdot \delta {\bm v}_i =0$.

After propagating up to the time of inverse Compton scattering, at time $t_{IC}$,
the velocity and position of one of the pair -- assumed positron -- is (the other member 
of the pair gets lost), 
\be
{\bm v}_{IC} = {\bm v}_i + {q} \frac{{\bm v}_i}{E} \times {\bm b},
\ee
\be
{\bm x}_{IC} = {\bm x}_i + {\bm v}_i T + {q} \frac{{\bm v}_i}{E} \times {\bm c},
\ee
where $E$ is the energy of the electron/positron, $q=+e$ is the positron's electric
charge, $T = t_{IC}-t_i$,
\be
{\bm b} = \int_{t_i}^{t_{IC}} dt'~ {\bm B} ({\bm x}_i + {\bm v}_i (t'-t_i)), 
\ee
\be
{\bm c} = \int_{t_i}^{t_{IC}} dt' \int_{t_i}^{t'} dt'' ~ {\bm B} ({\bm x}_i + {\bm v}_i (t''-t_i)) .
\label{cdefn}
\ee
Note that we are using the unperturbed trajectory when doing the integration. This
is legitimate since we are working to lowest order in $|{\bm B}|$.

The constraint is that the GeV photon should arrive to Earth. So 
\be
{\bm v}_{IC}=-{\hat {\bm x}}_{IC}.
\label{vICconstraint}
\ee
This gives
\be
{\bm v}_i + {q} \frac{{\bm v}_i}{E} \times {\bm b} = 
- \frac{1}{r_{IC}} \left ( {\bm x}_i + {\bm v}_i T + {q} \frac{{\bm v}_i}{E} \times {\bm c} \right ).
\ee
Ignoring terms of quadratic and higher order in $\delta{\bm v}_i$, ${\bm B}$, and also
assuming $T \ll r_{IC} \sim |{\bm x}_i| = r_i$ -- typically $T \sim 30~{\rm kpc}$, $r_i \sim 1~{\rm Gpc}$ --
gives
\be
\delta {\bm v}_i \approx {q} \frac{{\hat {\bm x}}_i}{E} \times \left ( {\bm b} + \frac{{\bm c}}{r_i} \right ).
\label{dvieq}
\ee
This is the constraint on the initial velocity so that the GeV gamma ray reaches Earth.
Straight-forward algebra now gives us the arrival direction of the photon
\begin{align}
{\bm n}(E) \equiv -{\bm v}_{IC} (E) =& -{\bm v}_i - {q} \frac{{\bm v}_i}{E}\times {\bm b} \nn \\
=& {\hat {\bm x}}_i - \delta{\bm v}_i - {q} \frac{{\bm v}_i}{E}\times {\bm b} \nn \\
\approx& {\hat {\bm x}}_i - {q} \frac{{\hat {\bm x}}_i}{E} \times \left ( {\bm b} + \frac{{\bm c}}{r_i} \right )
                 + {q} \frac{\hat{\bm x}_i}{E}\times {\bm b} \nn \\
=& {\hat {\bm x}}_i - {q} \frac{{\hat {\bm x}}_i}{E} \times \frac{\bm c}{r_i}.
\label{eq:ne}
\end{align}

Next we insert the expression for ${\bm n}$ in the definition of $Q_\infty$ in \eref{Qinfty}
and ignore linear and cubic terms in ${\bm B}$ to get
\be
Q_\infty =  \langle {\hat {\bm x}}_{i1} \times {\hat {\bm x}}_{i2} \cdot {\hat {\bm x}}_{i3} \rangle
 + \sum_{\rm cyc.} \frac{{q_1 q_2}}{E_e(E_1) E_e(E_2) r_{i1} r_{i2}} \langle
 ({\hat {\bm x}}_{i1} \times {\bm c}_1)\times ({\hat {\bm x}}_{i2} \times {\bm c}_2)
            \cdot {\hat {\bm x}}_{i3} \rangle,
            \label{Qformula}
\ee
where the $1,2,3$ subscripts refer to particles of energies $E_1$, $E_2$, $E_3$ 
and charges $q_1$, $q_2$, $q_3$; the sum is over cyclic perturbations of $1,2,3$,
and the subscript $i$ refers to the initial moment, at the time the TeV photon produces pairs. 
\eref{Qformula} is the expression for $Q_\infty$ to quadratic order in the magnetic field.

Note that the first term in \eref{Qformula} also depends on the magnetic field through
the constraint conditions \eref{xiconstraint} and \eref{vICconstraint}. In fact we shall show 
in a moment that the second term is much smaller than the first.
Focussing on the first term, using \eref{xiconstraint}, we can write
\be
\langle {\hat {\bm x}}_{i1} \times {\hat {\bm x}}_{i2} \cdot {\hat {\bm x}}_{i3} \rangle
 \approx \frac{1}{D_s^3} \langle ({\bm x}_s + D_{\rm TeV1} {\bm v}_{i1}) \times
                                                   ({\bm x}_s + D_{\rm TeV2} {\bm v}_{i2})\cdot
                                                   ({\bm x}_s + D_{\rm TeV3} {\bm v}_{i3} )\rangle,
\ee
where we have approximated $|{\bm x}_i| \approx D_s$. The initial velocities depend
on $\delta {\bm v}_i$ in \eref{vieq}, which in turn depend on the magnetic field via
\eref{dvieq}. Then, noting that $D_{\rm TeV}/D_s \ll 1$, we find
\be
\langle {\hat {\bm x}}_{i1} \times {\hat {\bm x}}_{i2} \cdot {\hat {\bm x}}_{i3} \rangle
\approx \sum_{(a,b)} \frac{D_{\rm TeV,a} D_{\rm TeV,b}}{D_s^2} 
           \langle \delta {\bm v}_{ia} \times \delta {\bm v}_{ib} \rangle \cdot {\hat {\bm x}}_s,
\ee
where the sum is over $(a,b)=(1,2),(2,3),(3,1)$. Now we can insert expressions for
$\delta {\bm v}_{ia}$ using \eref{dvieq}. The algebra simplifies because
\be
|{\bm b}| = O(D_e B) \gg \frac{|{\bf c}|}{r_i} = O\left ( \frac{D_e^2 B}{D_s} \right ),
\label{bcinequality}
\ee
where we recall $D_e \sim 30~{\rm kpc}$ and $D_s \sim {\rm Gpc}$. Then, to lowest
order in $D_e/D_s$ and $D_{\rm TeV}/D_s$ we obtain
\be
\langle {\hat {\bm x}}_{i1} \times {\hat {\bm x}}_{i2} \cdot {\hat {\bm x}}_{i3} \rangle
\approx \sum \frac{q_a q_b D_{\rm TeV,a} D_{\rm TeV,b}}{E_e(E_a) E_e (E_b) D_s^2} 
                      \langle  {\bm b}_a \times {\bm b}_b \rangle \cdot {\hat {\bm x}}_s.
                        \label{firsttermest}
\ee

Now we can justify our earlier claim that the first term in \eref{Qformula} dominates: using
the estimates in \eref{bcinequality}, the first term is of order $D_{\rm TeV}^2 D_e^2 B^2/D_s^2$,
whereas the second term is of order $D_{\rm e}^4 B^2/D_s^2$ which is smaller by the
factor $D_e^2/D_{\rm TeV}^2 \sim 10^{-6}$. Therefore
\be
Q_\infty \approx \sum \frac{q_a q_b D_{\rm TeV,a} D_{\rm TeV,b}}{E_{ea} E_{eb} D_s^2} 
                      \langle  {\bm b}_a \times {\bm b}_b \rangle \cdot {\hat {\bm x}}_s,
\ee
where we use the notation $E_{ea} = E_e(E_a)$.

\eref{magcorr} gives
\be
\langle  {\bm b}_a \times {\bm b}_b \rangle \cdot {\hat {\bm x}}_s
 = \int_{t_{ia}}^{t_{ICa}} dt' \int_{t_{ib}}^{t_{ICb}} dt'' ~ {\hat {\bm x}}_s \cdot ({\bm x}_a (t') - {\bm x}_b (t''))
                               M_H(|{\bm x}_a(t')-{\bm x}_b(t'')|),
 \label{babb}
\ee
where
\be
{\bm x}_a(t') = {\bm x}_a(t_i) + {\bm v}_{ia} (t' - t_{ia}),
\ee
and similarly for ${\bm x}_b(t'')$. Under the approximation that all deflection angles are small, we
get
\be
\langle  {\bm b}_a \times {\bm b}_b \rangle \cdot {\hat {\bm x}}_s
 \approx \int_{t_{ia}}^{t_{ICa}} dt' \int_{t_{ib}}^{t_{ICb}} dt'' d_{ab} M_H(|d_{ab}|),
\ee
where
\be
d_{ab} \equiv (r_{ia}-(t_a'-t_{ia})) - (r_{ib}-(t_b''-t_{ib})).
\ee

We can simplify further by using $r_{ia} \gg t_a''-t_{ia}$ ($a=1,2,3$), which
then implies $d_{ab} \approx r_{ia} - r_{ib}$ independent of the time integration variables.
If $(d_{ab}-(r_{ia}-r_{ib}))M_H'(d_{ab}) \ll M_H(d_{ab})$ where the prime on $M_H$ 
denotes differentiation with respect to the argument, then the integrand in 
\eref{babb} is approximately a constant and
\be
\langle  {\bm b}_a \times {\bm b}_b \rangle \cdot {\hat {\bm x}}_s
 \approx 2 D_{ea} D_{eb} (r_{ia}-r_{ib}) M_H (|r_{ia}-r_{ib}|) ,
\ee
where, $D_{ea} = D_e(E_a)= t_{ICa}-t_{ia}$, the time between pair production and IC scattering. 
To get a sense of numerical values, for the energies of 
interest, $D_{ea} \sim 30 {\rm kpc}$ and $r_{i1} \sim r_{i2} \sim {\rm Gpc}$, and 
$r_{i1}-r_{i2} \sim {\rm Mpc}$.

Finally we have the simple result
\be
Q_\infty \approx
\frac{2 e^2}{D_s^2}  \sum_{(a,b)}
\frac{q_{ab}D_{\rm TeV,a} D_{\rm TeV,b} D_{ea} D_{eb} }{E_{ea} E_{eb}}  d_{ab} M_H (|d_{ab}|),
\ee
where $q_{ab}=+1$ if $q_a=q_b$ and $q_{ab}=-1$ if $q_a \neq q_b$.

Note that 
\be
d_{ab} \approx r_{ia} - r_{ib} = (r_s - D_{\rm TeV,a}) - (r_s - D_{\rm TeV,b})
= D_{\rm TeV,b} - D_{\rm TeV,a} < 0,
\ee 
for $E_a < E_b$.

We now express $Q_\infty$ in terms of the observed energies. First we use
\eref{Deeq} that gives us the time between pair production and IC scattering, 
\be
D_e(E) \approx \frac{178 ~{\rm kpc}}{(1+z_s)^{4}} \left ( \frac{10~{\rm GeV}}{E} \right )^{1/2} 
 \equiv \left ( \frac{\delta_e}{E} \right )^{1/2},
 \label{deltae}
\ee
where 
$\delta_e \approx 3.2\times 10^5/(1+z_s)^8 ~ {\rm GeV\text{-}kpc^2} 
                      \approx 4.8\times 10^{34} /(1+z_s)^8~ {\rm Gpc}$
since we are working in natural units and we used $z_e \approx z_s$. 

Similarly
\be
 D_{\rm TeV} = \frac{237~{\rm Mpc}} {(1+z_s)^{2}} \left ( \frac{10~{\rm GeV}}{E} \right )^{1/2} 
\equiv \left ( \frac{\delta_T}{E} \right )^{1/2},
 \label{deltaT}
\ee
where $\delta_T \approx 5.6 \times 10^5/(1+z_s)^{4}~{\rm GeV\text{-}Mpc^2} 
\approx 8.8 \times 10^{40}/(1+z_s)^{4}~{\rm Gpc}$. We can assume $z_s \approx z_e$.
Also,
\be
E_e (E) \approx 1.7~{\rm TeV} \left(\frac{E}{10~\rm GeV}\right)^{1/2}
\equiv \left ( \frac{E}{\epsilon_e} \right )^{1/2},
\label{eq:energy-e}
\ee
with $\epsilon_e \approx 3.5\times 10^{-6}~{\rm GeV}^{-1}$.

Then,
\begin{align}
Q_\infty =& 
- \frac{{2 e^2} \delta_e \delta_T \epsilon_e}{D_s^2} \left [
 \frac{{q_{12}} |d_{12}| M_H (|d_{12}|)}{E_1^{3/2} E_2^{3/2}}
+\frac{{q_{23}} |d_{23}| M_H (|d_{23}|)}{E_2^{3/2} E_3^{3/2}}
-\frac{{q_{13}} |d_{13}| M_H (|d_{13}|)}{E_3^{3/2} E_1^{3/2}} 
\right ] \nonumber \\
=& 
- \frac{10^{27}}{(1+z_s)^{12} { \rm G^2}}  \left [
 \frac{{q_{12}} |d_{12}| M_H (|d_{12}|)}{{\cal E}_1^{3/2} {\cal E}_2^{3/2}}
+\frac{{q_{23}} |d_{23}| M_H (|d_{23}|)}{{\cal E}_2^{3/2} {\cal E}_3^{3/2}}
-\frac{{q_{13}} |d_{13}| M_H (|d_{13}|)}{{\cal E}_3^{3/2} {\cal E}_1^{3/2}} 
\right ] 
\left ( \frac{1~{\rm Gpc}}{D_s} \right )^2,
\label{Qinfty1}
\end{align}
where we have used $e^2 = 4\pi /137$, the Gauss to GeV conversion 
$1~{\rm G} = 1.95\times 10^{-20}~{\rm GeV}^2$, and denoted ${\cal E}_a = {E_a}/{10~{\rm GeV}}$.

Finally we can write the distances $d_{ab}$ in terms of the energies, since $d_{ab}$ is the
distance between the points where pair production occurs for photons of observed energy $E_a$
and $E_b$. Using \eref{eq:tev-meanfree} together with \eref{eq:gamma-ray},
\be
d_{ab} \approx r_{ia}-r_{ib} 
 = - \frac{237\kappa}{(1+z_s)^2} 
      \left ( \frac{1}{\sqrt{{\cal E}_a}} -  \frac{1}{\sqrt{{\cal E}_b}} \right ) ~{\rm Mpc}.
\label{eq:d_ab}
\ee

Eqs.~(\ref{Qinfty1}) and (\ref{eq:d_ab}) specify the correlator $Q_\infty$, {\it i.e.} the value
of $Q(R)$ in \eref{QR} for large values of $R$ in the case when only cascade photons are present.
The only required inputs are the helical magnetic correlation function and the gamma ray energies 
that are observed. Let us now consider $Q(R)$ for smaller values of
$R$, still only considering cascade photons. It  is useful to rewrite \eref{QR} as
\be
Q(R) = \langle ({\bm n} (E_1)-{\bm n}(E_3)) \times ({\bm n} (E_2)-{\bm n}(E_3)) \cdot {\bm n}(E_3) \rangle_R.
\ee
Then it is clear that $Q(R)$ depends on the length of the vector ${\bm n} (E_1)-{\bm n}(E_3)$ and
${\bm n} (E_2)-{\bm n}(E_3)$ where ${\bm n}(E_1)$ and ${\bm n}(E_2)$ are restricted to lie
in the patch of size $R$ with ${\bm n}(E_3)$ as center. The average lengths of these vectors are proportional
to $R$ for small $R$. Further, the cascade photons are clustered in the region around ${\bm n}_3$
and so the lengths do not grow indefinitely as $R$ increases. For $R$ larger than the typical bending 
angles of the 
cascade photons, the lengths can be taken to saturate. A functional form of $Q_c (R)$ that
takes the lengths of the vectors ${\bm n} (E_1)-{\bm n}(E_3)$ and ${\bm n} (E_2)-{\bm n}(E_3)$
into account in a patch of radius $R$ is
\be
Q_c (R) = (1-e^{-R/\Theta(E_1)})(1-e^{-R/\Theta(E_2)}) Q_\infty .
\label{QcR}
\ee

\subsection{Charge ambiguity}
\label{Z2ambiguity}

The formula for $Q_\infty$, \eref{Qinfty1}, includes the factors $q_{ab}=\pm 1$, that depend on
whether the cascade photons at energies $E_a$ and $E_b$ originated from like charges (electron-electron
or positron-positron) or unlike charges. The ambiguity can be traced back to the Lorentz force formula
that is invariant under charge reversal together with magnetic field reversal. We now discuss how
to possibly resolve the ambiguity due to the $q_{ab}$ factors in connecting observed $Q_c (R)$ to
the helical power spectrum $M_H$.

Let us assume that data are sufficiently precise that we can measure $Q_c(R, E_a,E_b,E_c)$ for
a wide range of energies. Then consider $Q_c(R,E_a,E_b,E_c)$ and $Q_c(R,E_a',E_b,E_c)$
where $E_a' \neq E_a$. By continuity, we expect 
$Q_c(R,E_a,E_b,E_c) \approx Q_c(R,E_a',E_b,E_c)$, and if this is indeed observed to be the case,
then $q_a = q_a'$. If, however, $Q_c(R,E_a,E_b,E_c)$ and $Q_c(R,E_a',E_b,E_c)$ are sharply
different, then $q_a = - q_a'$. In other words, to connect observed $Q_c$ to $M_H$ we need to
consider the different discrete possibilities and then choose the signs so as to reconstruct a continuous 
$M_H$\footnote{The discrete ambiguity is similar to the one that occurs when reconstructing magnetic fields
using Faraday Rotation measurements \citep{Zeldovichbook}.}.

The above resolution assumes very extensive observational data. Even without such data
but with the prior that the magnetic field is smooth on the scales of interest, we can take $q_{ab}=+1$
since without field reversals the cascades at different energies must arise from like charges. In what
follows, we will assume $q_{ab}=+1$.

\subsection{Source plus background}
\label{background}

The diffuse gamma ray sky contains cascade photons from many different TeV blazars, as
well as non-cascade photons from other sources. The expression for $Q_c (R)$ in
\eref{QcR} was derived for cascade gamma rays from a single TeV blazar, assuming no
contamination from other TeV blazars or non-cascade photons. We now extend our
calculation to include non-cascade photons.

The ensemble average over photons in some given patch of radius $R$ will include both
cascade, {\it i.e.} signal, and non-cascade, {\it i.e.} noise, photons. The value of $Q(R)$ will
depend on the ratio of cascade to non-cascade photons within that patch. The number of 
non-cascade photons within the patch will be proportional to the area of the patch
\be
N_{n}(E,R) = 2\pi \sigma_n (E) (1-\cos R) \equiv \sigma_n(E) A(R),
\ee
where $\sigma_n (E)$ is the average areal density of such ``noise'' photons with energy $E$,
and the area of a patch of radius $R$ on a unit sphere is $A(R) = 2\pi (1-\cos R)$.

The number of cascade photons within a patch of radius $R$ is also proportional to the
area of the patch when $R$ is less than the typical spread of cascade photons. For larger
$R$, the number of cascade photons within the patch stays constant. This suggests
\be
N_c (E) = N_\infty (E) (1 - e^{-A(R)/A(\Theta (E) )}),
\ee
where $N_\infty (E)$ is the number of cascade photons of energy $E$ when the patch size is
large, and $\Theta (E)$ is the typical bending angle at energy $E$. We stress that these
functional forms have the correct asymptotic dependencies but their precise shape should
not be taken too literally.
 
Then $Q(R)$, which is an average over all photons -- cascade and non-cascade -- will be 
given by
\be
Q(R) = 
\frac{N_{c}(E_1) N_{c}(E_2) N_c(E_3)}{N_{t}(E_1)N_{t}(E_2) N_t(E_3)} Q_c (R),
\label{QRQc}
\ee
where $N_{t} = N_{c} + N_{n}$. 

Note that in a given patch we only consider a single $E_3$
photon that is located at the center of the patch, but then $Q(R)$ is also averaged over all 
patches, some of which may be centered on $E_3$ photons that are not from a cascade.
This gives the energy $E_3$ a special status, and $N_c/N_t$ for $E_3$ is the average
number of cascade to non-cascade photons of energy $E_3$ over the region of sky under
consideration {\em e.g.} after the Milky Way has been masked out. So $N_c/N_t$ for $E_3$
is just a numerical factor; in particular, it does not depend on the radius of the patch. In
reconstructing the helical power spectrum from $Q(R)$, the value of $N_c(E_3)/N_t(E_3)$ 
will be a scaling factor. In what follows, we will denote 
\be
\frac{N_c(E_3)}{N_t(E_3)} = \frac{1}{1+\nu_3}
\label{nu3}
\ee
where $\nu_3$ is independent of $R$.

The ratio $N_c(E)/N_t(E)$ ($E=E_1, E_2$) depends on the radius of the patch and on the 
dimensionless ratio,
\be
\nu (E) \equiv \frac{\sigma_n (E) A(\Theta (E))}{N_\infty (E) (1-e^{-1})},
\ee
which is the ratio of the number of noise to the number of 
signal photons in a patch of angular radius $\Theta(E)$, 
{\it i.e.} the ``noise to signal ratio at energy $E$''.
Then,
\be
\frac{N_c}{N_t} = \left ( 1 + \frac{0.63~ \nu(E) {\cal A}(R, E)}{1- \exp(-{\cal A}(R,E))} \right )^{-1},
\ee
where ${\cal A}(R,E) = A(R)/A(\Theta (E)) = (1-\cos R)/(1-\cos \Theta(E))$.
Since the shape of $Q(R)$ is sensitive to $\nu (E)$, observations of $Q(R)$ may
in principle be used to determine $\nu (E)$ which would give a handle on the relative
number of cascade and non-cascade photons and also the number of TeV blazar sources. 
In the simplified analysis we present here, we will assume $\nu(E) \sim 1$ and
independent of the energy $E$.

\begin{figure}
\begin{center}
  \includegraphics[width=80mm]{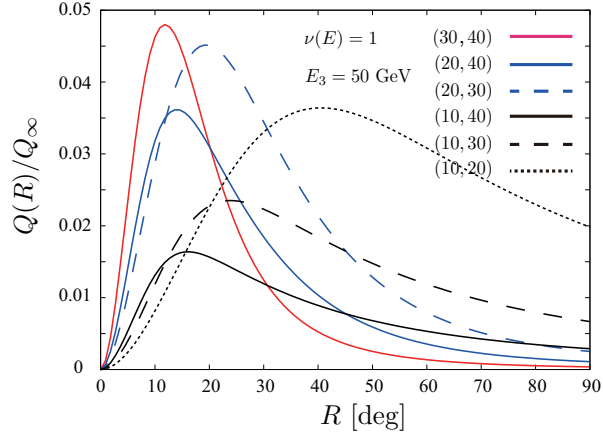}
  \end{center}
\caption{
Plot of $Q(R)/Q_\infty$ for $E_3=50~{\rm GeV}$ and $(E_1,E_2)$ as shown for $\nu(E) = 1$,
and with $B=2 \times 10^{-13}~{\rm G}$ and $D_s = 1~$Gpc to estimate $\Theta(E)$.
The lines in black, blue and red are for $E_1=10$, $20$ and $30$~GeV, respectively.
The dotted, dashed and solid lines are for $E_2=20$, 30 and 40~GeV. Note that
the peaks in the curves with the same $E_2$ are at approximately the same location.
}
\label{fig:QbyQinfty}
\end{figure}

In Fig.~\ref{fig:QbyQinfty} we plot $Q(R)/Q_\infty$ for sample parameters, showing that 
we expect magnetic helicity to lead to a peak in $Q(R)$. The location of the peak 
does not depend on the magnetic helicity power spectrum, $M_H$, but does depend on
the normal power spectrum, $M_N$, via the bending angles, $\Theta (E)$. 
The shape of the plots in Fig.~\ref{fig:QbyQinfty} can be understood as follows. Since
$E_1 < E_2$, and $\Theta (E) \propto E^{-3/2}$ from \eref{eq:final_ang_est}, we
have $\Theta(E_2) < \Theta(E_1)$. For $R  \ll \Theta(E_2)$ the
factors $N_c/N_t$ are approximately independent of $R$ and the exponential
factors in \eref{QcR} can be expanded to get $Q(R) \sim R^2$.
For $R \sim \Theta (E_2)$, $N_c/N_t$ at $E_1$ is still independent of $R$ since $R/\Theta(E_1)$
is small, but the $N_c/N_t$ factor for $E_2$ starts to vary with $R$ if
\be
\frac{0.63~ \nu(E_2) {\cal A}(R, E_2)}{1- \exp(-{\cal A}(R,E_2))}  > 1,\  {\rm and} , \ 
{\cal A}(R,E_2) > 1.
\label{e2conditions}
\ee
With $\nu(E) \approx 1$, these conditions are satisfied for ${\cal A}(R,E_2) \approx 1$.
%
Therefore $N_c/N_t$ for $E_2$ decreases as $\sim 1/R^2$ for $R > R_*$ where
${\cal A}(R_*, E_2) \approx 1$
which gives
\be
R_* \approx \Theta (E_2).
\ee
Thus the shape of $Q(R)$ is flat for $R \sim R_*$.
For $\Theta(E_2) < R < \Theta (E_1)$, similarly we find $Q(R) \sim 1/R$, and finally
for $\Theta (E_1) < R$, we get $Q(R) \sim 1/R^4$. 

The important features of the shape of $Q(R)$ are that it has a peak located at
\be
R_{\rm peak} \approx R_* \approx  \Theta(E_2)
\label{Rpeak}
\ee
and the peak height is found by substituting $R=R_{\rm peak}$ in \eref{QRQc} 
and using Eqs.~(\ref{QcR}) and (\ref{Qinfty1}),
\be
Q_{\rm peak} \approx
\frac{0.63}{(1+0.63\nu_1)(1+\nu_2)(1+\nu_3)}
\frac{\Theta(E_2)}{\Theta (E_1)} Q_\infty 
\approx
0.1 \left ( \frac{E_1}{E_2} \right )^{3/2} Q_\infty
\label{QpeakQinfty}
\ee
where $\nu_i  = \nu(E_i)$, with $\nu_i =1$ in the final estimate, and
we have assumed $\Theta(E_2) \ll \Theta(E_1)$.
The amplitude of the peak depends on $Q_\infty$, 
which depends on the magnetic helicity spectrum, $M_H$.
In Sec.~\ref{samplemodels} we will consider some models of the helicity power spectra and plot $Q(R)$.
In Sec,~\ref{conclusions} we will use these peak characteristics to estimate the intergalactic
magnetic field strength.

\section{Sample models of $M_H$}
\label{samplemodels}

\begin{figure}
  \begin{center}
	\includegraphics[width=80 mm]{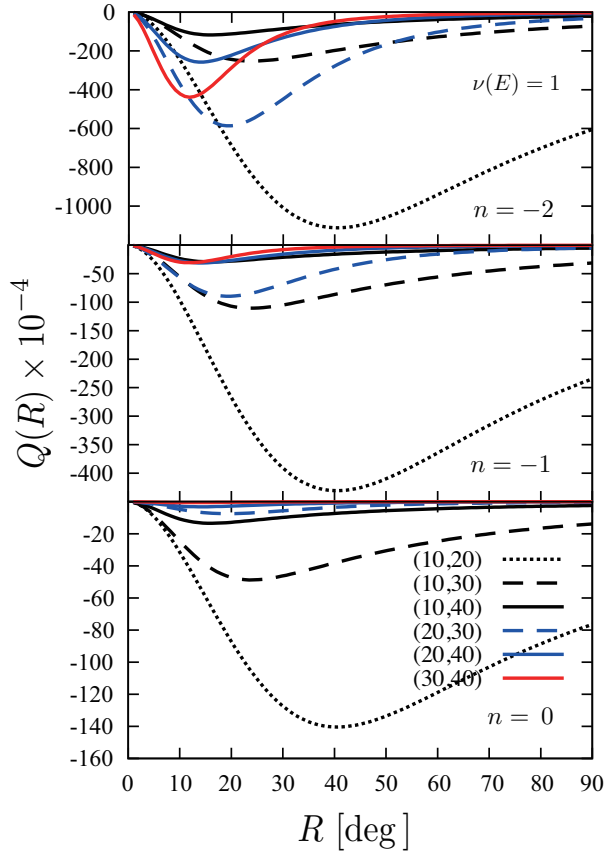}
  \end{center}
  \caption{
 Plot of $Q(R)$ for different spectral index $n$ of $M_H$. We set $\nu(E) = 1$. From the top to bottom panels, the spectral index
 is set to $n=-2$, -1 and 0, respectively. 
The energy combinations are represented with the same types of lines
 as in Fig.~\ref{fig:QbyQinfty}.}
\label{fig:QvsR}
\end{figure}

We now calculate $Q(R)$ from Eq.~(\ref{QRQc}) with $Q_c$ given by \eref{QcR} and $Q_\infty$ by 
\eref{Qinfty1} with $q_{ab}=1$ (see Sec.~\ref{Z2ambiguity}).
For simplicity, we assume a power-law form for the helical power spectrum,
\begin{equation}
 r M_H(r) = B_N^2 \frac{r}{r_N} \left(\frac{|r|}{r_N} \right)^n, 
 \label{MHpowerlaw}
\end{equation}
where $B_N$ is the normalized magnetic field strength at the normalization scale $r_N$. 
In our context, $B_N$ is defined at a redshift $z_s$ since that is roughly where the 
cascade is being produced. The conversion to
the comoving magnetic field strength $B_0$ is $B_0 = B_N/(1+z_s)^2$.
We adopt $B_N = 2\times 10^{-13}~{\rm G}$ and $r_N = 100~{\rm Mpc}$.
According to Eq.~(\ref{magcorr}), $n = -1$ represents a scale-invariant 
helicity spectrum, $n>-1$ gives 
a red-tilted spectrum, and $n<-1$ gives a blue-tilted spectrum.
The choice of $B_N$ is made to match the peak scale of the energy
combination (10,40) as seen in the Fermi-LAT data~($R_* \approx 14^\circ$)
in \citet{Tashiro:2013ita}. The plots of $Q(R)$ versus $R$ are shown in 
Fig.~\ref{fig:QvsR}.


Whereas we have adopted the form in \eref{MHpowerlaw} and predicted $Q(R)$, we can also reverse
these arguments and deduce $M_H(r)$ from observational data.

\section{Conclusions}
\label{conclusions}

In this paper we have related the statistics, $Q(R)$, calculated from the diffuse gamma ray
sky, to the helical power spectrum, $M_H$, of intergalactic magnetic fields. Our results
show that helical intergalactic magnetic fields lead to a peak structure in $Q(R)$. The
angular location of the peak determines the typical bending angle of cascade photons of a
given energy, and is in turn related to the non-helical power spectrum, $M_N$, of
the magnetic field; the amplitude of the peak is related to the helical power spectrum,
$M_H$. The sign of $Q$ depends on the handedness of the magnetic field and, except 
when there is a lot of power on small distance scales, we expect $Q < 0$ for left-handed fields. 
In addition, the shape of $Q(R)$ is sensitive to the fraction of cascade to non-cascade photons
in the sky, thus possibly providing a tool to study the distribution and properties of 
cascade photons from TeV blazars.

Using our results, we can obtain a numerical estimate of the helical power spectrum.
From \eref{Qinfty1} we estimate
\be
\left | \, d_{12} M_H(|d_{12}|) \, \right | \sim  
(3 \times 10^{-16}~ {\rm G})^2 {\cal E}_1^{3/2} {\cal E}_2^{3/2} 
 \left ( \frac{D_s}{1~{\rm Gpc}} \right )^2 \left ( \frac{Q_\infty}{10^{-4}} \right ) (1+z_s)^{12}
 \sim 
\left [ 1.5 \times 10^{-14} D_{s1} (1+z_s)^6 ~ {\rm G} \right ]^2 ,
\label{magestimate}
\ee
where, in the last estimate we have used $E_1 = 10~{\rm GeV}$,  $E_2 = 40~{\rm GeV}$,
{\it i.e.} ${\cal E}_1=1$, ${\cal E}_2 =4$, $Q_{\rm peak} = 3\times 10^{-4}$ \citep{Tashiro:2013ita},
$Q_\infty = 80~ Q_{\rm peak}$ from \eref{QpeakQinfty} with $\nu(E_i)=1$,
and denoted $D_{s1} = D_s/1~{\rm Gpc}$. Therefore we estimate
\be
B_0 \approx 1.5 \times 10^{-14} D_{s1} (1+z_s)^4 ~ {\rm G} , \ \ \ {\rm from~peak~amplitude}
\label{B0frompeakamplitude}
\ee
where $B_0$ is the magnetic field strength at the present epoch and we have taken the
field strength to evolve in proportion to $(1+z)^2$.
From~\citet{Tashiro:2013ita} we find that the peak in $Q(R)$ is located at $\Theta_s \approx 14^\circ$
when $E_1 = 10~{\rm GeV}$,  $E_2 = 40~{\rm GeV}$.
Then from \eref{eq:final_ang_est}, with $E_\gamma = 40~{\rm GeV}$ -- recall
that the peak position is determined by $E_2$ as in \eref{Rpeak} -- we find
\be
B_0 \approx 8 \times 10^{-14} D_{s1} (1+z_s)^4~ {\rm G}, \ \ \  {\rm from~peak~location}
\label{B0frompeaklocation}
\ee
This rough agreement between the two independent estimates in \eref{B0frompeakamplitude}
and \eref{B0frompeaklocation} is quite remarkable and suggests that the intergalactic magnetic 
fields are significantly, if not maximally, helical.

In conclusion, we have developed the connection between our proposed parity odd correlator,
$Q(R)$, and properties of the intergalactic magnetic field. 
We find encouraging agreement between independent estimates 
of the magnetic field strength using observed data, thus providing further confidence in the 
robustness of the signal discussed in~\citet{Tashiro:2013ita}.
There is also some tentative indication in the observed $Q(R)$ to peak at smaller
angles at higher $E_2$ \citep{Tashiro:2013ita} which is in agreement with the trend in Fig.~\ref{fig:QvsR}.
Another interesting observation is that if we combine \eref{QpeakQinfty} and \eref{Qinfty1} and
assume that the first term in \eref{Qinfty1} dominates, we find
\be
Q_{\rm peak} \sim
-  \frac{10^{26}}{(1+z_s)^{12} { \rm G^2}} 
 \frac{{q_{12}} |d_{12}| M_H (|d_{12}|)}{{\cal E}_2^{3/2}}
\left ( \frac{1~{\rm Gpc}}{D_s} \right )^2 .
\ee
Hence, for a scale invariant spectrum, $Q_{\rm peak}$ depends only on $E_2$ and there is
little dependence on $E_1$. We have also seen that $R_{\rm peak}$ is independent
of $E_1$ (see \eref{Rpeak}). Thus both $R_{\rm peak}$ and $Q_{\rm peak}$ are
only sensitive to $E_2$ for a spectrum that is not too steep. Indeed, in the
plots obtained from Fermi data for $E_2=40~{\rm GeV}$ there is a peak with amplitude
$\sim 3 \times 10^{-4}$ at roughly $R=14^\circ$ regardless of the value of $E_1$, again
showing consistency with the intergalactic magnetic fields hypothesis. Further, 
\eref{Rpeak} predicts that $Q(R)$ will peak at $R \approx 21^\circ$ when $E_2=30~{\rm GeV}$
and at $R \approx 40^\circ$ when $E_2=20~{\rm GeV}$; the height of the peak is sensitive
to the noise to signal ratio, $\nu(E)$, and to the helicity power spectrum.

We close with a few cautionary remarks. Our analytic methods have been possible only because we 
have made several simplifying assumptions along the way. For example, we have ignored 
the stochasticity of the pair production -- which depends on the EBL spectrum --  and the 
CMB up-scattering processes. Monte Carlo methods seem to be most suitable for including these 
processes though
simulations will be challenging because of the range of length scales involved (kpc to Gpc). We hope
the rewards of discovering helical intergalactic magnetic fields and the enormous implications for particle
physics and cosmology will spur further investigations.

\section*{acknowledgements}
We are grateful to Wenlei Chen, Francesc Ferrer, and Andrew Long for comments.
TV thanks IAS, Princeton for hospitality while this work was being done.
This work was supported by MEXT's Program for Leading Graduate Schools ``PhD
 professional: Gateway to Success in Frontier Asia,''
the Japan Society for Promotion of Science (JSPS) Grant-in-Aid for Scientiffic Research
(No.~25287057) and the DOE at ASU.


\appendix

\section{Sign of $Q$, $M_H$ and magnetic handedness}
\label{sec:sign}
Define
\be
C_{ij} = \frac{1}{2} \langle B_i ({\bm x}+{\bm r}) B_j ({\bm x}) - B_j ({\bm x}+{\bm r}) B_i ({\bm x}) \rangle 
         = \epsilon_{ijl}r^l M_H(r).
\ee
Then
\be
r_k \epsilon^{kij} C_{ij} = 2 r^2 M_H(r).
\ee
Let us evaluate the left-hand side directly for a right-handed magnetic field configuration. Take ${\bm x}=0$,
${\bm r} = r {\hat {\bm z}}$ ($r > 0$), ${\bm B}({\bm x}=0) = B {\hat {\bm x}}$, and
${\bm B}({\bm x}= r {\hat {\bm z}}) = B {\hat {\bm y}}$. This is a right-handed configuration because
the magnetic field rotates counterclockwise as one goes from the origin to $r {\hat {\bm z}}$.
Then
\be
2 r^2 M_H(r) = r_k \epsilon^{kij} C_{ij} =
r {\hat {\bm z}} \cdot {\bm B}( r {\hat {\bm z}}) \times {\bm B} (0) = 
r B^2 {\hat {\bm z}} \cdot {\hat {\bm y}} \times {\hat {\bm x}} = - r B^2 < 0.
\ee
So we find, with our conventions, $M_H(r) < 0$ for a right-handed magnetic field configuration.

From \eref{Qinfty1}, assuming that the lowest energy photon term dominates,
\be
Q \sim 
- \frac{{2 e^2} \delta_e \delta_T \epsilon_e}{D_s^2}
 \frac{{q_{ab}} |d_{ab}| M_H (|d_{ab}|)}{E_a^{3/2} E_b^{3/2}}.
\ee
Therefore, if $q_{ab} = +1$ as for smooth fields, $Q < 0$ implies $M_H > 0$ which implies 
left-handed magnetic helicity with our conventions. 
This conclusion is reversed if $q_{ab}=-1$ or if the helical spectrum is such that the last term in 
\eref{Qinfty1} dominates, which could happen if there is significant power on small scales.


\end{document}